\documentclass[letterpaper,twocolumn,aps,prl,floatfix,superscriptaddress]{revtex4-1}

\usepackage{float}	
\usepackage[caption=false]{subfig}
\usepackage{graphicx}	
\usepackage{url}
\usepackage{dsfont}
\usepackage{physics}
\usepackage{breqn}
\usepackage{siunitx}
\usepackage[colorlinks=true]{hyperref}
\usepackage[capitalise]{cleveref}
\crefname{section}{Sec.}{Secs.}
\Crefname{section}{Section}{Sections}
\makeatother

\makeatletter
\let\cat@comma@active\@empty
\makeatother

\graphicspath{ {./figs/}}

\newcommand{\ahat}{\hat{a}}

\newcommand{\Hhat}{\hat{\mathcal{H}}}
\newcommand{\Vhat}{\hat{V}}

\newcommand{\dhat}{\hat{d}}
\newcommand{\alphahat}{\hat{\alpha}}
\newcommand{\chat}{\hat{c}}
\newcommand{\bhat}{\hat{b}}
\newcommand{\yhat}{\hat{y}}
\newcommand{\zhat}{\hat{z}}
\newcommand{\xhat}{\hat{x}}

\newcommand{\Shat}{\hat{S}}
\newcommand{\Phat}{\hat{P}}
\newcommand{\Qhat}{\hat{Q}}
\newcommand{\done}{\hat{d}^{(1)}}
\newcommand{\dtwo}{\hat{d}^{(2)}}
\newcommand{\radialint}{\mathcal{R}}

\newcommand{\idop}{\mathds{1}}

\newcommand{\Deltap}[1]{\Delta_+^{(#1)}} 
\newcommand{\Deltam}[1]{\Delta_-^{(#1)}} 
\newcommand{\ketrp}[1]{\ket*{r_+^{(#1)}}}
\newcommand{\ketrm}[1]{\ket*{r_-^{(#1)}}}
\newcommand{\brarp}[1]{\bra*{r_+^{(#1)}}}
\newcommand{\brarm}[1]{\bra*{r_-^{(#1)}}}
\newcommand{\ketgp}[1]{\ket*{g_+^{(#1)}}}
\newcommand{\ketgm}[1]{\ket*{g_-^{(#1)}}}

\newcommand{\Csix}[1]{C_6^{(#1)}}

\newcommand{\Omegap}[1]{\Omega_+^{(#1)}}
\newcommand{\Omegam}[1]{\Omega_-^{(#1)}} 

\newcommand{\khz}{\si{\kHz}}
\newcommand{\Mhz}{\si{\MHz}}
\newcommand{\Ghz}{\si{\GHz}}
\newcommand{\mum}{\si{\um}}
\newcommand{\mics}{\si{\micro\second}}
\begin{document}	
\title{Nondestructive cooling of an atomic quantum register via state-insensitive Rydberg interactions}
\date{\today}
\author{Ron Belyansky}
\affiliation{Joint Quantum Institute, NIST/University of Maryland, College Park, Maryland 20742 USA}
\author{Jeremy T. Young}
\affiliation{Joint Quantum Institute, NIST/University of Maryland, College Park, Maryland 20742 USA}
\author{Przemyslaw Bienias}
\affiliation{Joint Quantum Institute, NIST/University of Maryland, College Park, Maryland 20742 USA}
\author{Zachary Eldredge}
\affiliation{Joint Quantum Institute, NIST/University of Maryland, College Park, Maryland 20742 USA}
\affiliation{Joint Center for Quantum Information and Computer Science, NIST/University of Maryland, College Park, Maryland 20742, USA}
\author{Adam M. Kaufman}
\affiliation{JILA, University of Colorado and National Institute of Standards and Technology, and Department of Physics, University of Colorado, Boulder, Colorado 80309, USA}
\author{Peter Zoller}
\affiliation{Institute for Quantum Optics and Quantum Information, Austrian Academy of Sciences \& Center for Quantum Physics, University of Innsbruck, Innsbruck A-6020, Austria}
\author{Alexey V. Gorshkov}
\affiliation{Joint Quantum Institute, NIST/University of Maryland, College Park, Maryland 20742 USA}
\affiliation{Joint Center for Quantum Information and Computer Science, NIST/University of Maryland, College Park, Maryland 20742, USA}

\begin{abstract}
We propose a protocol for sympathetically cooling neutral atoms without destroying the quantum information stored in their internal states. This is achieved by designing state-insensitive Rydberg interactions between the data-carrying atoms and cold auxiliary atoms.
The resulting interactions give rise to an effective phonon coupling, which leads to the transfer of heat from the data atoms to the auxiliary atoms, where the latter can be cooled by conventional methods.  
This can be used to extend the lifetime of quantum storage based on neutral atoms and can have applications for long quantum computations.
The protocol can also be modified to realize state-insensitive interactions between the data and the auxiliary atoms but tunable and non-trivial interactions among the data atoms, allowing one to simultaneously cool and simulate a quantum spin-model. 
\end{abstract}
\maketitle
\pagenumbering{arabic}
{\it Introduction.---}%
In recent years, neutral atoms stored in individual traps have emerged as a powerful resource for quantum information and quantum technologies. 
Individual atoms can be trapped in optical \cite{Grimm2000} or magnetic \cite{Boetes2018,Fortagh2007} potentials forming arrays of various geometries \cite{Barredo2018,Endres2016,Kim2016}, providing a platform for quantum simulators \cite{Bernien2017,Labuhn2016,Lienhard2017,Zeiher2017} and quantum computers \cite{Saffman2018,Saffman2016}.
Their long-lived ground states can be used to store quantum information and the highly excited Rydberg states can be used to implement strong and tunable interactions \cite{Jaksch2000,Saffman2016,Glaetzle2015,VanBijnen2015,Browaeys2016}.

Considerable effort is currently being invested in developing neutral atom traps that are insensitive to the internal state of the atom \cite{Zhang2011,Ye2008,Topcu2016,Boetes2018}.
These so-called magic traps attempt to achieve what is naturally available with trapped ions, since the trapping of the latter relies on the net charge of the ion, and hence is independent of its internal electronic state. 
The magic trapping of neutral atoms reduces heating and dephasing associated with the fact that different electronic states may have different trapping potentials.
Nevertheless, even with such magic trapping conditions, heating of the motional degrees of freedom of the atoms can occur because of, for example, the shaking of the atomic array due to laser intensity noise \cite{Savard1997}, mechanical forces from Rydberg interactions \cite{Jaksch2000,Saffman2005,Saffman2010}, or incoherent light scattering \cite{Pichler2010}.

Such heating of the atomic motion, when combined with state-dependent Rydberg mediated gates, generally leads to reduced fidelities and loss of coherence, which is particularly problematic for long quantum simulations or computations \cite{Saffman2011,Kumar2018,Wang2016}.
It is therefore desirable to develop schemes to cool the atomic motion without destroying the quantum information stored in the internal states.
The conventional laser cooling techniques \cite{Thompson2013,Kaufman2012,Sompet2017} are not suitable for this task since they involve optical pumping which, in general, destroys the quantum information.

Several approaches for this problem have already been proposed in the past, from immersing the atomic lattice in a superfluid \cite{Daley2004} to using cavity-assisted cooling \cite{Griessner2004}. 
It has also been shown that alkaline-earth atoms can be laser-cooled without destroying the quantum information provided it is stored in the nuclear spin \cite{Reichenbach2007}.

In this paper, we introduce two schemes of achieving state-insensitive interactions between neutral atoms, another natural and useful tool of trapped ions. We further show how to use these interactions to realize a state preserving cooling procedure, inspired by sympathetic cooling of trapped ions \cite{Barrett2003,Kielpinski2002}.
In contrast to the protocols in Refs. \cite{Daley2004,Griessner2004}, ours requires only ingredients and capabilities that are already present in many neutral atoms experiments: auxiliary atoms and Rydberg interactions.

\begin{figure}[htb]
	\centering
	\includegraphics[width=\linewidth]{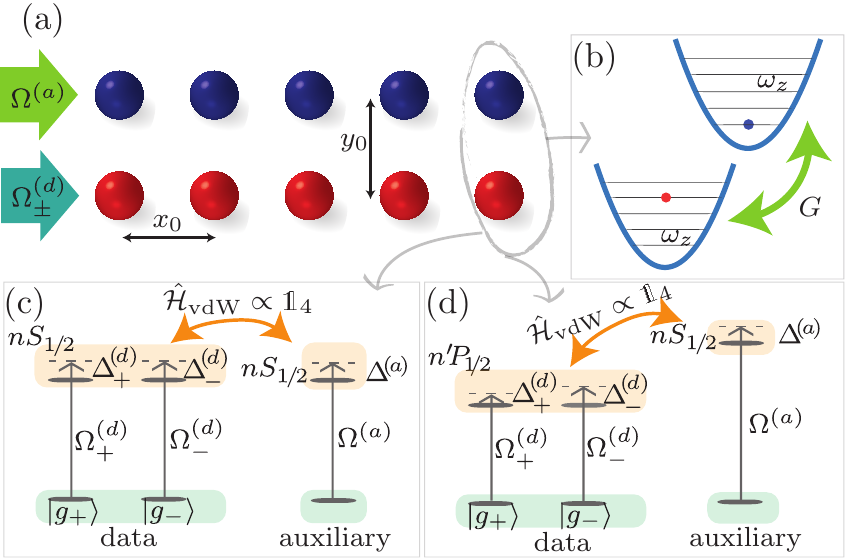}
	\caption{Schematic of the phonon-swap protocol. (a) For each data atom (red, bottom) we place another auxiliary atom (blue, top) at an equal distance $y_0$. Here we assume a 1D chain of data and auxiliary atoms, with a lattice spacing of $x_0$. (b) The Rydberg interactions give rise to effective coupling $G$ between the vibrational modes of the data and auxiliary atoms. (c,d) Two possible schemes that lead to spin-insensitive interactions between the data and the auxiliary atoms: in (c), the spin states (represented by the ground states $\ket{g_+},\ket{g_-}$) of the data atoms, as well as the ground state of the auxiliary atoms, are all weakly coupled to highly-excited Rydberg $S_{1/2}$ states. In (d), the data atoms are coupled to $n'P_{1/2}$ states and the auxiliary to $nS_{1/2}$, where $\abs{n-n'}\gg1$. This leads to spin-independent interactions between the data and auxiliary atoms, but tunable spin-spin interactions among the data atoms.} 
	\label{fig:diagram}
\end{figure}

The scenario we have in mind is the following: we assume one starts with a quantum data register composed of an array of $N$ atoms, each in an individual trap, cooled to the vibrational ground state and optically pumped to a particular ground state. Each atom encodes useful information in its ground states in the form of a two-level system which we represent by a spin-1/2. 
One then uses Rydberg interactions to perform a quantum computation or a quantum simulation, during which the atoms are heated, as mentioned above. To cool the data register we introduce $N$ additional auxiliary atoms, one for each data atom [see for example \cref{fig:diagram}(a)], that have been precooled using any of the standard methods.

The data and auxiliary atoms can then be coupled via Rydberg interactions, implementing a \emph{phonon-swap} gate -- a coherent exchange of vibrational quanta between the data and auxiliary atoms.

Since the phonon-swap is mediated by Rydberg interactions, a key requirement for this protocol is for the interactions between the auxiliary and data atoms to be insensitive to the internal state of the data atoms. 
Unlike the Coulomb interaction between trapped ions which naturally satisfies this requirement, the Rydberg interactions between neutral atoms are inherently state-dependent. As we show in this Letter, a careful choice of the Rydberg states can nevertheless lead to state-insensitive data-auxiliary interactions.

Another requirement for this protocol is that these data-auxiliary interactions used to generate the phonon-swap should not induce unwanted state-dependent couplings between the data atoms.
We present two schemes [see \cref{fig:diagram}(c,d) respectively], both of which satisfy the two requirements but lead to different interactions amongst the data register atoms. 
In the first scheme, the interactions between any pair of atoms (data-data and data-auxiliary) are independent of the internal state. This scheme therefore consists of interrupting the quantum computation or simulation, performing the phonon-swap, and then resuming the computation or simulation.
In the second scheme, the data and auxiliary atoms are addressed separately, and this allows one to design the interactions in such a way that the data-auxiliary interactions are state-insensitive but the data-data interactions are tunable and controllable. As an example, we show how this can be used to implement the phonon-swap while simultaneously performing a quantum simulation of a spin model on the data atoms.
Finally, for both of the above schemes that we discuss, one can laser cool the auxiliary atoms during the phonon-swap. 
Due to the quantum Zeno effect \cite{Itano1990}, this has the additional advantage of preventing certain coherent heating mechanisms, such as those due to the Rydberg interactions themselves,from taking place in the first place.
We leave the detailed study of such a scheme for future work.

{\it Phonon-swap for two atoms.---}%
To illustrate the phonon-swap mechanism, let us first consider the case of two atoms: one two-level data atom ``$d$" and another single-level auxiliary atom ``$a$". The two atoms are each trapped in a three-dimensional harmonic potential separated by a distance $\vb{r}$. 
In many recent experiments \cite{AlexandreCooperJacobCoveyIvayloMadjarov,Kaufman2012,Yu2018,Barredo2018,Thompson2013,Sompet2017}, the confinement along two directions ($x,y$) is often much stronger than along the third ($z$), i.e $\omega_x,\omega_y\gg \omega_z$, where $\omega_\alpha$ is the trap frequency along the direction $\alpha\in \{x,y,z\}$.
Here, for simplicity, we therefore focus on cooling a single trap component, which we take to be the weakest direction ($z$). Cooling the two components that are perpendicular to the inter-atomic axis of the two atoms is a trivial generalization of what we present in this section. The third component, which is along the inter-atomic axis [$y$ axis in \cref{fig:diagram}(a)], requires more care but can be cooled via an adiabatic protocol, the details of which are presented in the Supplemental Material \cite{sup}.

The Hamiltonian consisting of both the vibrational and the internal degrees of freedom is ($\hbar=1$ throughout) 
$\Hhat=  \omega_{z}(\dhat^\dagger\dhat+\ahat^\dagger\ahat)+\Hhat_{s}+\Hhat_{\text{int}}(\vb{r})$,
where $\dhat\,(\ahat)$ is the phonon annihilation operator of the data (auxiliary) atom along the $z$ direction; $\Hhat_{s}$ is a Hamiltonian that acts on the internal (spin) degree of freedom of the data atom, and $\Hhat_{\text{int}}(\vb{r})$ is the interaction Hamiltonian between the two atoms that, in principle, couples motion and spin. 
Since we do not want to affect the spin state of the data atom, we need to decouple the phonon dynamics from the spin. In other words, we want $\Hhat_{\text{int}}=\idop_{\text{internal}}\otimes V(r)$ to be an identity operator on the internal states. 
As we later show, by weakly laser-dressing the ground states with Rydberg states, it is possible to obtain effective interactions of such form, where the spatial dependence is $V(r)=\frac{\mathcal{A}}{r^6+R_c^6}$ for some coupling $\mathcal{A}$ and blockade radius $R_c$, irrespective of the spin state.

For now, let us assume these interactions and Taylor-expand them to second order in the small quantum fluctuations on top of the macroscopic separation $r_0$, which we assume to be along one of the strongly confined directions [see for example \cref{fig:diagram}(a)]. This gives rise to a quadratic Hamiltonian in terms of the bosonic phonon-annihilation operators of the two atoms $\ahat,\dhat$ \cite{sup},
\begin{dmath}
	\label{eq:phonon-ham-2}
	\Hhat_{\text{ph},2}=\omega_z(\dhat^\dagger\dhat+\ahat^\dagger\ahat)-\frac{G}{2}\qty[(\dhat+\dhat^\dagger)^2+(\ahat+\ahat^\dagger)^2]+G(\dhat+\dhat^\dagger)(\ahat+\ahat^\dagger),
\end{dmath}
where $G= \frac{3\mathcal{A}}{M\omega_zr_0^{8}}\frac{1}{[1+(R_c/r_0)^6]^2}$ is the effective phonon coupling strength and $M$ is the mass of each atom.
In the regime where $\omega_{z}\gg G$, only the number-conserving terms are relevant, giving a ``beam splitter" interaction (in the rotating frame) 
$\Hhat_{\text{ph},2} = G(\ahat^\dagger\dhat+\dhat^\dagger\ahat)$.
This Hamiltonian effectuates a state-transfer between the two vibrational modes in a time of $t_s = \frac{\pi}{2G}$, swapping in the process the phonons of the data atom with those of the auxiliary atom. This effectively cools the data atom down to the initial phonon occupancy of the auxiliary atom, provided that the latter is initially colder.

\begin{figure}[tb]
	\centering
	\includegraphics[width=\linewidth]{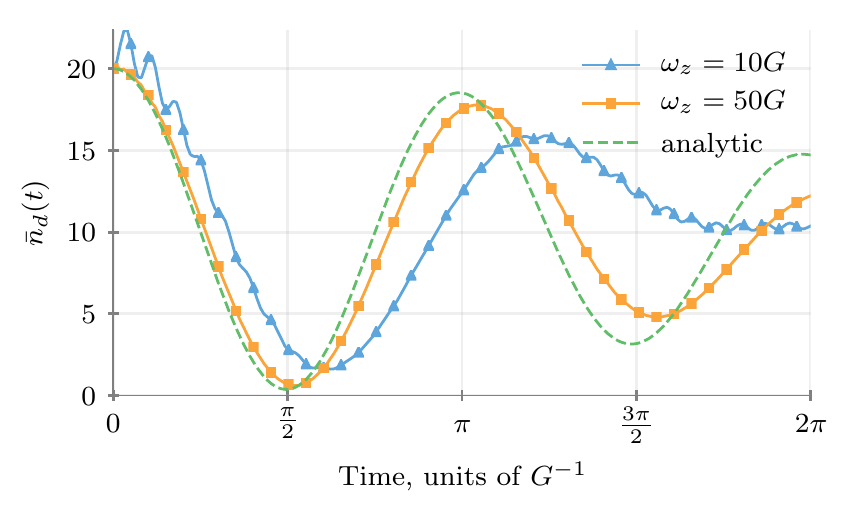}
	\caption{The average number of phonons in the data atoms as a function of time (in units of $G^{-1}$) computed numerically (solid lines) for different values of $\omega_z$ for two chains of 50 atoms, including the counter-rotating terms in \cref{eq:phonon-ham-2}, and analytically (dashed line) using the approximation of \cref{eq:average_n_data}. Here, $\eta=1$ and the initial conditions are $\bar{n}_d(0)=20$ and $\bar{n}_a(0)=0$.} 
	\label{fig:n_data}
\end{figure}

{\it Phonon-swap for 1D chain.---}%
The discussed protocol can be easily generalized for an ensemble of atoms. We simply associate with each data atom we would like to cool a cold auxiliary atom. 
For concreteness, we consider a chain of data atoms with a lattice constant $x_0$, brought to a distance of $y_0$ from a chain of cold auxiliary atoms, as shown in \cref{fig:diagram}(a). In the same regime as above, the many-body Hamiltonian is quadratic with approximate power-law decaying  hopping between the sites \cite{sup}
\begin{equation}
	\begin{split}
		\Hhat_{\text{ph},1D}&=\sum_{i\neq j}G_{ij}(\ahat_i^\dagger\ahat_j+\dhat_i^\dagger\dhat_j)+\sum_{ij}F_{ij}(\ahat_i\dhat_j^\dagger+\ahat_i^\dagger\dhat_j),\\
		G_{ij} &=\frac{G}{\eta^{8}\abs{i-j}^{8}},\label{eq:H-ph-1D}\\
		F_{ij} &=\frac{G}{\qty[\eta^2(i-j)^2+1]^{4}},
	\end{split}
	\raisetag{3\normalbaselineskip}
\end{equation}
where $\eta \equiv \frac{x_0}{y_0}$. Here we defined $G$ in terms of the smallest distance between a data atom and its auxiliary, i.e $y_0$ (see \cref{fig:diagram}).
Clearly, as $\eta\rightarrow \infty$, it is sufficient to consider only the nearest-neighbor interactions between data and auxiliary atoms. In such case, we should recover the situation discussed in the previous paragraph, namely each data-auxiliary pair perfectly swaps their phonons after a time of $t_s=\frac{\pi}{2G}$. If we also take into account next-nearest-neighbor interactions between the data and auxiliary atoms, we find \cite{sup} that the average phonon occupancy of the data atoms is given by
\begin{dmath}
	\label{eq:average_n_data}
	\bar{n}_d(t)=\frac{\bar{n}_a(0)+\bar{n}_d(0)}{2}  -\frac{\bar{n}_a(0)-\bar{n}_d(0)}{2}J_0\qty[\frac{4Gt}{(1+\eta^2)^4}]\cos(2Gt),
\end{dmath}
where $\bar{n}_{d}(t)$ ($\bar{n}_{a}(t)$) is the average occupancy of data (auxiliary) atoms at time $t$ and $J_0(z)$ is a Bessel function of the first kind.
\Cref{eq:average_n_data} is quantitatively accurate (see \cref{fig:n_data}) at short time-scales, when the effects of the long range interactions are less important.
As $\eta \rightarrow \infty$, $J_0\rightarrow 1$ and we reproduce the case of independent pair-wise phonon-swaps.
Moreover, as can be seen in \cref{fig:n_data}, $t_s = \frac{\pi}{2G}$ is still the nearly optimal swap time and even with $\eta=1$ we can still achieve a high-efficiency swap. 
Assuming for simplicity that the auxiliary atoms are initially in the vibrational ground state, we obtain a swap efficiency of $1-\frac{\bar{n}_d(t_s)}{\bar{n}_d(0)}= \frac{1}{2}+\frac{1}{2}J_0(\frac{\pi}{8})\approx 98\%$.
Furthermore, \cref{eq:average_n_data} remains qualitatively accurate even at longer time-scales. As $t\rightarrow \infty$, $J_0 \rightarrow0$ and we see that the mean phonon occupancy of all atoms is the average of the total initial number of phonons, as one would expect.

The conclusion of the above discussion is that in order to cool an atomic register consisting of many atoms in arbitrary geometries and dimensions, we simply perform the phonon-swap as if all the data-auxiliary pairs are independent. 
The effects of the many-body interactions only lead to a small degradation in the swap efficiency. 

{\it State-insensitive Rydberg interactions.---}%
We now turn to discuss how to obtain the spin-independent interactions by utilizing the strong van-der-Waals (vdW) coupling between highly-excited Rydberg states. 
Specifically, we concentrate on alkali atoms and consider weakly laser-admixing two hyperfine ground states (see Supplemental Material for an explicit example \cite{sup}) representing the spin-1/2, $\ket{g_+},\ket{g_-}$, to Rydberg states $\ket{r_+},\ket{r_-}$, depicting the magnetic sublevels of either $S_{1/2}$ or $P_{1/2}$ manifolds, as shown in \cref{fig:diagram}(c,d). 
The vdW couplings $\Hhat_{\text{vdW}}$ between the Rydberg states then get imprinted onto the ground states, giving effective interactions between the dressed ground states.
The relevant Hamiltonian describing this is 
$\Hhat = \sum_{i=1,2}(\Hhat_A^{(i)}+\Hhat_L^{(i)})+\Hhat_{\text{vdW}}$ where $\Hhat_A^{(i)}=-\Deltap{i}\ketrp{i}\brarp{i}-\Deltam{i}\ketrm{i}\brarm{i}$ and $\Hhat_L^{(i)} = \frac{\Omegap{i}}{2}\ketgp{i}\brarp{i} +\frac{\Omegam{i}}{2}\ketgm{i}\brarm{i}+\text{H.c.}$ are the atomic and laser Hamiltonians, respectively, in the rotating frame after applying the rotating wave approximation. Here, $\Omega_{\pm}^{(i)}$ are the two Rabi frequencies and $\Delta_{\pm}^{(i)}\gg \Omega_{\pm}^{(i)}$ the laser detunings.
Note that for the auxiliary atoms, it is sufficient to consider a single ground state and hence a single laser. However, we must take into account all the states in the Rydberg manifold. 
This is because, in general, $\Hhat_{\text{vdW}}$ can contain both diagonal and off-diagonal matrix elements which can mix all the states in the Rydberg manifold. This fact has been used previously to construct complex, tunable spin-spin interactions \cite{Glaetzle2015,VanBijnen2015}.
A sufficient condition to obtain spin-independent interactions is, therefore, for $\Hhat_{\text{vdW}}$ to be proportional to an identity, together with a suitable choice of the laser parameters.
We show below two simple schemes using $S_{1/2}$ and $P_{1/2}$ states that yield $\Hhat_{\text{vdW}} \propto \idop$ to a good approximation.

The vdW Hamiltonian between two atoms in either $S_{1/2}+S_{1/2}$, $S_{1/2}+P_{1/2}$ or $P_{1/2}+P_{1/2}$ in the Zeeman basis (i.e $\{\ket{++},\ket{+-},\ket{-+},\ket{--}\}$) has the following form \cite{sup}
\begin{align}
	\Hhat_{\text{vdW}}&=
	\frac{C_6}{r^6}\idop_4-\frac{C_6^{(a)}+C_6^{(b)}-C_6^{(c)}-C_6^{(d)}}{r^6}\mathcal{D}_0(\theta,\phi),\label{eq:Hvdw-final}\raisetag{3\normalbaselineskip}\\ 
	C_6&=\frac{2}{27}\qty[C_6^{(a)}+4C_6^{(b)}+2(C_6^{(c)}+C_6^{(d)})],\label{eq:C6}
\end{align}
where the $C_6^{(p)}$ coefficients correspond to the four different channels describing the possible $(L,J)$ quantum numbers of the intermediate states and $\mathcal{D}_0(\theta,\phi)$ is a traceless matrix that depends on the orientation of the interatomic axis relative to the quantization axis. The channels for $S_{1/2}+S_{1/2}$ and $S_{1/2}+P_{1/2}$ are shown in \cref{tab:SS-and-SP-channels}.

\begin{table}[tb]
	\begin{ruledtabular}
		\begin{tabular}{ccc}
			& $S_{1/2} + S_{1/2}$ & $S_{1/2} + P_{1/2}$ \\ 
			$\begin{aligned}(a)\\
			(b)\\
			(c)\\
			(d)\end{aligned}$& $\begin{aligned} &S_{1/2}+S_{1/2} \rightarrow P_{1/2}+ P_{1/2}\\
			&S_{1/2}+S_{1/2} \rightarrow P_{3/2} + P_{3/2}\\
			&S_{1/2}+S_{1/2} \rightarrow P_{3/2} + P_{1/2}\\
			&S_{1/2}+S_{1/2} \rightarrow P_{1/2} + P_{3/2} \end{aligned}$&
			$\begin{aligned}		 &S_{1/2}+P_{1/2} \rightarrow P_{1/2}+ S_{1/2}\\
			&S_{1/2}+P_{1/2} \rightarrow P_{3/2} + D_{3/2}\\
			&S_{1/2}+P_{1/2} \rightarrow P_{3/2} + S_{1/2}\\
			&S_{1/2}+P_{1/2} \rightarrow P_{1/2} + D_{3/2}\end{aligned}$\\
		\end{tabular} 
		\caption{The four channels describing the dipole-allowed virtual processes $(L_1,J_1)+(L_2,J_2)\rightarrow (L_1',J_1')+(L_2',J_2')$ that lead to vdW interactions: (left) in the case of both atoms in $S_{1/2}$ states; (right) in the case of one atom in $S_{1/2}$ and the other in $P_{1/2}$.}
		\label{tab:SS-and-SP-channels}
	\end{ruledtabular}
\end{table}

{\it Phonon-swap with $S+S$ states.---}%
The first scheme uses the well known fact that for the case of a pair of atoms in $nS_{1/2}$ states, the second term in \cref{eq:Hvdw-final} approximately vanishes \cite{Walker2008,Glaetzle2015}. This can be seen from \cref{tab:SS-and-SP-channels}, which shows that the difference between the four channels is only in the fine structure of the intermediate states. In the limit of vanishing fine structure, we then have $\Csix{a}=\Csix{b}=\Csix{c}=\Csix{d}$. 
This can also be understood intuitively as follows:
the vdW interactions arise from second-order perturbation theory, where the two electrons undergo virtual transitions to intermediate states allowed by the dipole selection rules.
If we neglect the fine structure, we are free to use the uncoupled basis ($\ket{L,m_L}\otimes\ket{S,m_S}$) for the intermediate levels. Since $S_{1/2}$-states are proportional to definite electronic spin states with definite $m_S$, i.e $\ket{S_{1/2},m_J=\pm \frac{1}{2}}=\ket{L=0,m_L=0}\otimes \ket{S=\frac{1}{2},m_S=\pm \frac{1}{2}}$
and because the dipole-dipole interactions do not act on the electronic spin, the vdW couplings cannot mix states with different $m_J$.
The correction to this scales as $\Delta_{\text{FS}}/\delta$, where $\Delta_{\text{FS}}$ is the fine-structure splitting and $\delta$ the energy difference to the intermediate states.

Neglecting these small corrections, and to fourth order in the small parameter $\epsilon=\Omega/2\Delta$, the effective spin-spin interactions between any two data atoms are given by
\begin{dmath}
	\label{eq:dressed-diag-H-int}
	\Hhat_{\text{int}}(r) = \mqty(\dmat[0]{\tilde{V}_{++},\tilde{V}_{+-},\tilde{V}_{-+},\tilde{V}_{--}}).
\end{dmath}
In the case of data-auxiliary interactions,  we have a $2\times2$ version of \cref{eq:dressed-diag-H-int}. In both cases, the matrix elements are given by
\begin{dmath}
	\label{eq:V-dressed-munu}
	\tilde{V}_{\mu\nu}=\qty(\frac{\Omega_{\mu}^{(1)}\Omega_{\nu}^{(2)}}{4\Delta^{(1)}_{\mu}\Delta^{(2)}_{\nu}})^2\frac{C_6}{r^6-\frac{C_6}{\Delta^{(1)}_{\mu}+\Delta^{(1)}_{\nu}}},
\end{dmath}
which are spin-independent (i.e $\tilde{V}_{++} = \tilde{V}_{+-}=\tilde{V}_{-+}=\tilde{V}_{--}$) for a suitable choice of the laser parameters. A trivial example consists of the two laser fields being identical.
\begin{figure}[tb]
	\centering
	\includegraphics[width=\linewidth]{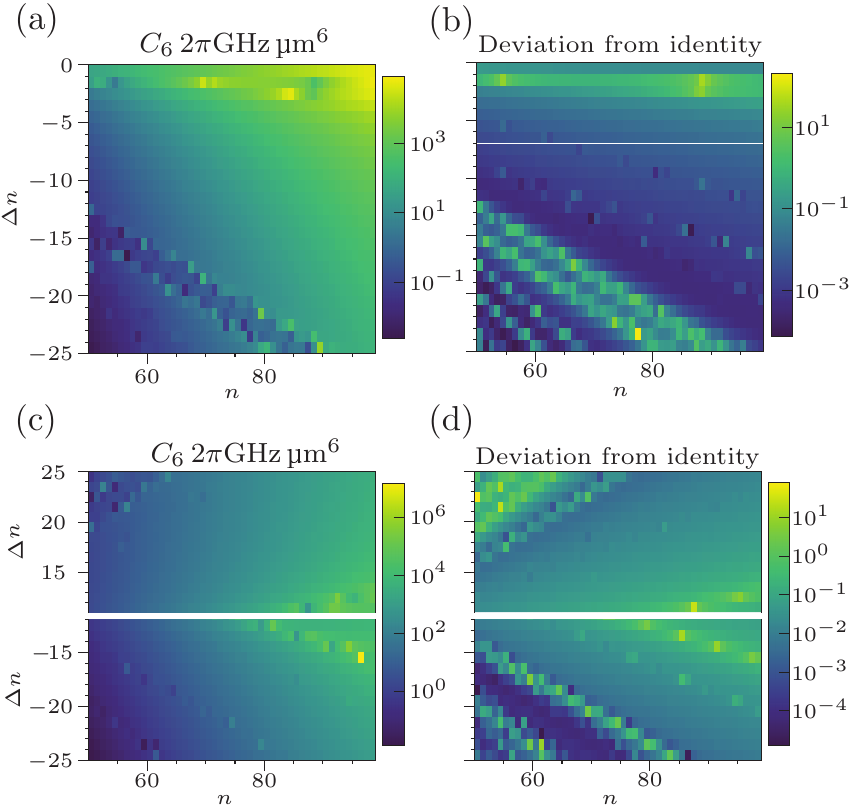}
	\caption{(a,c) The spin-insensitive interaction strength $C_6$ and (b,d) deviation from identity for (top) $nS_{1/2}+n'S_{1/2}$ and (bottom) $nS_{1/2} + n'P_{1/2}$ as a function of $n$ and $\Delta n = n'-n$ for Rb atoms. In the case of $nS_{1/2} + n'P_{1/2}$, we take $\min{\abs{\Delta n}}=10$ in order for the direct, dipolar coupling between the states to be negligible.
	} 
	\label{fig:C6}
\end{figure}
The cooling protocol with this scheme would thus consist of stopping the quantum simulation or computation, weakly coupling the ground states of both the data and auxiliary atoms to $nS_{1/2}$ states, and waiting for a time of $t_s$.  
As an example, Rb atoms separated by $2.36\,\mum$, and weakly coupled to $60S_{1/2}$ ($C_6/2\pi \approx 138.5\,\Ghz\,\mum^6$) with $\Omega/2\pi = 100\,\Mhz$\footnote{This can be achieved by a two-photon transition with one of the lasers tightly focused through an objective, or by using a build-up cavity. One can also use stronger interactions together with optimal control techniques to reduce the Rabi frequency.} and $\Delta/2\pi = 200\,\Mhz$ would experience a phonon coupling of $G/2\pi \approx  1.48 \,\khz$ assuming a trap frequency of $\omega_z/2\pi = 15\,\khz$. $G$ is about an order of magnitude smaller than the trap frequency and about two orders of magnitude larger than the effective decay rate $\epsilon^2\Gamma_{60S}/2\pi \approx 0.043\, \khz$, where $\Gamma_{60S}$ is the decay rate of $60S_{1/2}$ states.
The deviation of $\Hhat_{\text{vdW}}$ from identity, which we define by the ratio of the operator norms of the two terms in \cref{eq:Hvdw-final}, is in this case $\sim0.027$. 
This error can be reduced by driving the two atoms to different principal quantum numbers, as can be seen in panel (b) of \cref{fig:C6}. This generally reduces the $C_6$ coefficient, as can be seen in panel (a) of \cref{fig:C6}, but it can nevertheless be sufficiently strong with the additional advantage of reducing the magnitude of the spin-dependent couplings. For instance, $74S_{1/2}+64S_{1/2}$ yield $C_6/2\pi\approx 29\,\Ghz\,\mum^6$ (only a factor of five smaller than for $60S_{1/2}+60S_{1/2}$) with an error of $\sim0.003$ (an order of magnitude smaller).

{\it Phonon-swap with $S+P$ states.---}%
This brings us to the second scheme, in which the auxiliary atoms are coupled to $nS_{1/2}$ states, while the data atoms to $n'P_{1/2}$ states, where $\abs{\Delta n} = \abs{n'-n}\gg 1$ in order to ensure that the dipolar interactions between them can be ignored. This is because for large $\abs{\Delta n}$, the overlap of the wavefunctions and consequently the matrix element are small \cite{Gras2018}.
Such a configuration not only gives spin-independent interactions between the data and auxiliary atoms, as we will explain below, but also gives rise to non-trivial, tunable spin-spin interactions between the data atoms \cite{Glaetzle2015}. Thus, this allows one to cool the data atoms while simultaneously performing a quantum simulation of a spin model, for example.
To see why $S_{1/2}+P_{1/2}$ gives rise to $\Hhat_{\text{vdW}}\propto \idop$, note that channels $(a,c)$, as well as $(b,d)$, in \cref{tab:SS-and-SP-channels} only differ by the fine structure in one of the terms. In the limit of vanishing fine structure, the four channels cancel each other pair-wise, eliminating  $\mathcal{D}_0(\theta,\phi)$ in \cref{eq:Hvdw-final}.
\begin{figure}[t]
	\centering
	\includegraphics[width=\linewidth]{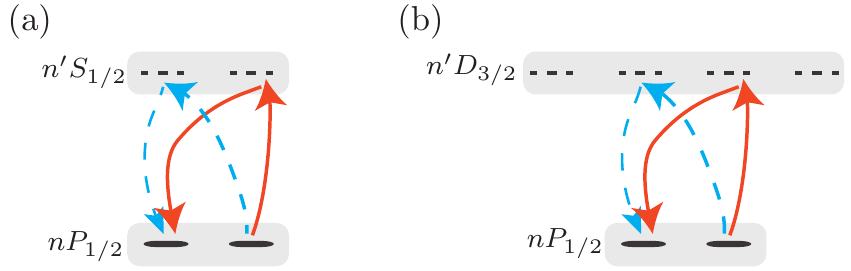}
	\caption{The four virtual transitions that can couple the $m_J=1/2$ magnetic state to $m_J=-1/2$, in the $P_{1/2}$ manifold. The second atom is assumed to be in a $S_{1/2}$ state and we are neglecting the fine structure in its intermediate manifold. The red (solid) and blue (dashed) transitions for each of the two possible sub-channels, (a) $P_{1/2}\rightarrow S_{1/2}$ and (b) $P_{1/2}\rightarrow D_{3/2}$, destructively interfere.} 
	\label{fig:P-virtual-trans}
\end{figure}
A more intuitive argument uses the same reasoning that was used to argue that vdW interactions between $S_{1/2}+S_{1/2}$ cannot couple states with different $m_J$. Similarly, in this case, there cannot be any mixing between states involving different $m_J$ of the $S_{1/2}$ atom.
Hence, in the absence of fine structure in the intermediate manifold, the $S_{1/2}$ atom is effectively decoupled and $\Hhat_{\text{vdW}}$ must at least be block-diagonal.  
Within this approximation, we can understand why the remaining off-diagonal matrix elements also vanish by focusing solely on the $P_{1/2}$ atom.
As can be seen in \cref{fig:P-virtual-trans}, for each possible sub-channel of the $P_{1/2}$ atom, there are exactly two processes that can, in principle, couple its $m_J=+\frac{1}{2}$ and $m_J=-\frac{1}{2}$ states. These two processes, however, precisely destructively interfere. 

The resulting dressed spin interactions between the data and auxiliary atoms then have the same form as in \cref{eq:dressed-diag-H-int,eq:V-dressed-munu}. The corresponding $C_6$, as well as the error due to spin-dependent couplings, can be seen in panels c and d of \cref{fig:C6}, respectively.

The data atoms, on the other hand, experience non-trivial spin-spin interactions due to the  $P_{1/2}+P_{1/2}$ vdW couplings. For the configuration in \cref{fig:diagram} (quantization axis parallel to interatomic axis), the $\mathcal{D}_0(\theta,\phi)$ matrix from \cref{eq:Hvdw-final} has the following form
\begin{dmath}
	\label{eq:D0-theta-zero}
	\mathcal{D}_0(0,\phi)=\frac{2}{81}\left(
	\begin{array}{cccc}
		1 & 0 & 0 & 0 \\
		0 & -1 & -4 & 0 \\
		0 & -4 & -1 & 0 \\
		0 & 0 & 0 & 1 \\
	\end{array}
	\right),
\end{dmath}
which gives rise to the following spin-1/2 Hamiltonian for the data atoms:
\begin{dmath}
	\Hhat =\sum_{ij}J_z^{ij}\Shat_z^{(i)}+J_{zz}^{ij}\Shat_z\Shat_z+\qty(J_{+-}^{ij}\Shat_+^{(i)}\Shat_-^{(j)}+\text{H.c.}),
\end{dmath}
where $\Shat_\alpha^{(i)}$ are the spin-1/2 operators of atom $i$ and $J_{\mu\nu}^{ij}$ are coefficients that depend on the geometry, laser parameters, and Rydberg interactions \cite{Glaetzle2015}.
This approach can be extended to generate other spin-1/2 models, for instance in two dimensions \cite{Glaetzle2015}, with simultaneous cooling.

{\it Summary and outlook.---}%
We have presented a protocol for sympathetically cooling Rydberg atoms without destroying the quantum information stored in their internal states. This can have applications for future Rydberg-based quantum computers and quantum simulators as well as other quantum technologies. 
We note that while we focused here on the weak coupling regime ($G\ll\omega_{z}$), which inevitably limits the phonon-swap time to $\sim1/G\gg 1/\omega_z$, it is possible to speed it up by working in the strong coupling regime $G\sim \omega_z$ and employing optimal control techniques \cite{MacHnes2010,Wang2011,MacHnes2012}. 
Finally, our schemes of realizing state-insensitive interactions between neutral atoms could also be used in other contexts, such as generating interesting states combining motional \cite{Buchmann2017} and electronic degrees of freedom. %

R.B., J.T.Y., P.B., Z.E., and A.V.G. acknowledge support by AFOSR, ARL CDQI, NSF PFC at JQI, the DoE ASCR Quantum Testbed Pathfinder Program (award No. DE-SC0019040), DoE BES QIS program (award No. DE-SC0019449), NSF PFCQC program, and ARO MURI.
R.B. acknowledges in addition support of NSERC and FRQNT. Z.E. is supported in part by the ARCS foundation. A.M.K. acknowledges support by NIST. P.Z. was supported by PASQuanS EU Quantum Flagship.

\bibliography{refs}

\onecolumngrid
\clearpage
\newpage 
\setcounter{figure}{0}
\makeatletter
\renewcommand{\thefigure}{S\@arabic\c@figure}
\setcounter{equation}{0} \makeatletter
\renewcommand \theequation{S\@arabic\c@equation}
\setcounter{table}{0} \makeatletter
\renewcommand \thetable{S\@arabic\c@table}

\begin{center}
	\textbf{\large Supplemental Materials: Nondestructive cooling of an atomic quantum register via state-insensitive Rydberg interactions}
\end{center}
This supplemental material is organized as follows: in \cref{sec:vdw-derivation-app}, we derive the Hamiltonian for the van-der-Waals interactions between the Zeeman sublevels of two atoms in either $S_{1/2}+S_{1/2}$, $S_{1/2}+P_{1/2}$, or $P_{1/2}+P_{1/2}$. In \cref{sec:phonon-interactions-app}, we derive the Hamiltonian for the phonon interactions between two atoms and discuss how to implement the phonon-swap for up to two trap components. 
We also comment on the validity of the Taylor approximation.
Then, in \cref{sec:adiabatic_cooling-app}, we present the adiabatic protocol for the phonon-swap and discuss how to perform 3D cooling by swapping all three trap components.
In \cref{sec:1dchain-deriv-app}, we generalize the phonon-swap for a 1D chain of data and auxiliary atoms and derive the time-dependence of the average phonon number in each species.
Finally, in \cref{sec:laser-dressing-states-app}, we give an example of the spin-1/2 states and the choice of the laser polarizations for a $^{87}$Rb atom.

\section{vdW interactions}
\label{sec:vdw-derivation-app}

In this section, we derive the van-der-Waals interactions (\cref{eq:Hvdw-final,eq:C6,eq:D0-theta-zero} in the main text) between the Zeeman sublevels of two atoms.
In second order perturbation theory, this can be written as \cite{Glaetzle2015}
\begin{equation}
	\label{eq:Hvdw-def}
	\Hhat_{\text{vdW}} = \Phat \sum_{\alpha,\beta}\frac{\Vhat_{dd}\Qhat_{\alpha,\beta}\Vhat_{dd}}{\delta_{\alpha\beta}}\Phat,
\end{equation}
where $\Qhat_{\alpha,\beta} = \ket{\alpha,\beta}\bra{\alpha,\beta}$ and $\Phat = \sum_{k,l}\ket{k,l}\bra{k,l}$ are projectors onto the intermediate and initial states, respectively.
The dipole-dipole operator, $\Vhat_{dd}$, is given by
\begin{equation}
	\label{eq:Vdd}
	\Vhat_{dd} = -\sqrt{\frac{24\pi}{5}}\frac{1}{r^3}\sum_{\mu,\nu}C_{\mu,\nu;\mu+\nu}^{1,1;2}Y_2^{\mu+\nu}(\theta,\phi)^*\done_\mu\dtwo_\nu,
\end{equation}
where $C_{m_1,m_2;M}^{J_1,J_2;J}$ is a Clebsch-Gordan coefficient and $Y_l^m$ spherical harmonics. $\done_\mu$ and $\dtwo_\nu$ are the spherical components of the dipole operators for the two atoms ($\mu, \nu \in \{-1,0,1\}$), whose matrix elements are 
\begin{equation}
	\label{eq:dipole-op-matrix-elem}
	\bra{n_a,L_a,J_a,m_a}\hat{d}_q\ket{n_c,L_c,J_c,m_c} \equiv\radialint_{n_a,L_a,J_a,n_c,L_c,J_c}J^q_{L_a,J_a,m_a,L_c,J_c,m_c},
\end{equation}
where
\begin{equation}
	\label{eq:radial-overlap}
	\radialint_{n_a,L_a,J_a,n_c,L_c,J_c}=\int R_{n_a,L_a,J_c}(r)R_{n_c,L_c,J_c}(r)r^3 d r,
\end{equation}
is the overlap of the radial wavefunctions $R_{n,L,J}(r)$ and
\begin{equation}
	\label{eq:Jangular-defs}
	\begin{split}
		J^q_{L_a,J_a,m_a,L_c,J_c,m_c} &=(-1)^{2J_c+1/2+m_a}\sqrt{(2J_a+1)(2J_c+1)(2L_a+1)(2L_c+1)}\\ &\mqty(J_c&1&J_a\\m_c&q&-m_a)\mqty(L_a&1&L_c\\0&0&0)\begin{Bmatrix}J_a&1&J_c\\L_c&1/2&L_a\end{Bmatrix}.
	\end{split}
\end{equation}

We can write the sum over the intermediate states $\alpha,\beta$ in \cref{eq:Hvdw-def} as follows
\begin{equation}
	\label{eq:interm-sum}
	\sum_{\alpha,\beta} = \sum_{\text{channels}}\sum_{n_{\alpha},n_{\beta}}\sum_{m_{\alpha},m_{\beta}},
\end{equation}
where the channels for $S_{1/2}+S_{1/2}$, $S_{1/2}+P_{1/2}$ and $P_{1/2}+P_{1/2}$  are given in \cref{tab:SS-SP-PP-channels}.

\begin{table}[htb]
	\begin{ruledtabular}
		\begin{tabular}{cccc}
			& $S_{1/2} + S_{1/2}$ & $S_{1/2} + P_{1/2}$ &  $P_{1/2} + P_{1/2}$ \\ 
			$\begin{aligned}(a)\\
			(b)\\
			(c)\\
			(d)\end{aligned}$& $\begin{aligned} &S_{1/2}+S_{1/2} \rightarrow P_{1/2}+ P_{1/2}\\
			&S_{1/2}+S_{1/2} \rightarrow P_{3/2} + P_{3/2}\\
			&S_{1/2}+S_{1/2} \rightarrow P_{3/2} + P_{1/2}\\
			&S_{1/2}+S_{1/2} \rightarrow P_{1/2} + P_{3/2} \end{aligned}$&
			$\begin{aligned}		 &S_{1/2}+P_{1/2} \rightarrow P_{1/2}+ S_{1/2}\\
			&S_{1/2}+P_{1/2} \rightarrow P_{3/2} + D_{3/2}\\
			&S_{1/2}+P_{1/2} \rightarrow P_{3/2} + S_{1/2}\\
			&S_{1/2}+P_{1/2} \rightarrow P_{1/2} + D_{3/2}\end{aligned}$& $\begin{aligned} &P_{1/2}+P_{1/2} \rightarrow S_{1/2}+ S_{1/2}\\
			&P_{1/2}+P_{1/2} \rightarrow D_{3/2} + D_{3/2}\\
			&P_{1/2}+P_{1/2} \rightarrow S_{1/2} + D_{3/2}\\
			&P_{1/2}+P_{1/2} \rightarrow D_{3/2} + S_{1/2} \end{aligned}$\\
		\end{tabular} 
		\caption{The four channels describing the dipole-allowed virtual processes $(L_1,J_1)+(L_2,J_2)\rightarrow (L_1',J_1')+(L_2',J_2')$ that lead to vdW interactions, in the case of both atoms in $S_{1/2}$ states (left), one atom in $S_{1/2}$ and the other in $P_{1/2}$ (middle), both atoms in $P_{1/2}$ states (right).}
		\label{tab:SS-SP-PP-channels}
	\end{ruledtabular}
\end{table}

Note that specifying the channel specifies both the $L$ and $J$ quantum numbers (the principal quantum numbers $n$ are also implicitly specified for the left-hand-side of the channel).

Using \cref{eq:interm-sum,eq:dipole-op-matrix-elem,eq:Vdd} we can rewrite \cref{eq:Hvdw-def} in the following form
\begin{align}
	\label{eq:HvdW-long}
	\begin{split}
		\Hhat_{\text{vdW}} =& \frac{1}{r^6}\sum_{\text{channels}}\qty(\sum_{n_{\alpha},n_{\beta}}\frac{(\radialint_{\alpha,1})^2(\radialint_{\beta,2})^2}{\delta_{\alpha\beta}})\times\\
		&\sum_{\substack{m_k,m_l \\ m_k',m_l'}}\qty[\frac{24\pi}{5}\sum_{m_{\alpha},m_{\beta}}\qty(\sum_{\substack{\mu,\nu \\ \mu',\nu'}}C_{\mu,\nu;\mu+\nu}^{1,1;2}Y_2^{\mu+\nu \,*}J_{1_k,\alpha}^{\mu}J_{2_l,\beta}^{\nu} C_{\mu',\nu';\mu'+\nu'}^{1,1;2}Y_2^{\mu'+\nu'\,*}J_{\alpha,1_{k'}}^{\mu'}J_{\beta,2_{l'}}^{\nu'})\ket{m_k,m_l}\bra{m_{k'},m_{l'}}].
	\end{split}
\end{align}
In \cref{eq:HvdW-long}, each term in the parentheses on the first line only depends on the intermediate $n_\alpha,n_\beta$ values, for a given channel. The label $\alpha$ in $\radialint_{\alpha,1}$ is short for ${n_\alpha,L_\alpha,J_\alpha}$ where $L_\alpha,J_\alpha$ are specified by the channel. Similarly, the label $1$ ($2$) is specifying the ${n,L,J}$ values of the first (second) term in the channel.

The quantity in the second line of \cref{eq:HvdW-long} is a $4\times4$ matrix in the subspace of the magnetic sublevels $\ket{++},\ket{+-},\ket{-+},\ket{--}$. For a given channel, the matrix elements are found by summing over the $m_\alpha,m_\beta$ values and are independent of $n$.

Thus, for a given channel, $p$, we can define a $C_6^{(p)}$ coefficient and a matrix $\mathcal{D}^{(p)}$
\begin{align}
	\label{eq:c6-Dmatrix-def}
	\begin{split}
		C_6^{(p)} =& \sum_{n_{\alpha},n_{\beta}}\frac{(\radialint_{\alpha,1})^2(\radialint_{\beta,2})^2}{\delta_{\alpha\beta}},\\
		\mathcal{D}^{(p)}_{kl,k'l'} =& \frac{24\pi}{5}\sum_{m_{\alpha},m_{\beta}}\qty(\sum_{\substack{\mu,\nu \\ \mu',\nu'}}C_{\mu,\nu;\mu+\nu}^{1,1;2}Y_2^{\mu+\nu \,*}J_{1_k,\alpha}^{\mu}J_{2_l,\beta}^{\nu} C_{\mu',\nu';\mu'+\nu'}^{1,1;2}Y_2^{\mu'+\nu'\,*}J_{\alpha,1_{k'}}^{\mu'}J_{\beta,2_{l'}}^{\nu'}).
	\end{split}
\end{align}
With these, the van-der-Waals Hamiltonian takes the simple form
\begin{equation}
	\label{eq:Hvdw-simple}
	\Hhat_{\text{vdW}} = \frac{1}{r^6}\sum_{p}C_6^{(p)}\mathcal{D}^{(p)}.
\end{equation}

For the channels in \cref{tab:SS-SP-PP-channels} (same results for all three cases) we find (different definition than in \cite{Glaetzle2015})
\begin{align}
	\begin{split}
		\mathcal{D}^{(a)} &= \frac{2}{27}\idop - \mathcal{D}_0,\\
		\mathcal{D}^{(b)} &= \frac{8}{27}\idop - \mathcal{D}_0,\\
		\mathcal{D}^{(c)} &= \frac{4}{27}\idop + \mathcal{D}_0,\\
		\mathcal{D}^{(d)} &= \frac{4}{27}\idop +\mathcal{D}_0,
	\end{split}
\end{align}
where 
\begin{equation}
	\mathcal{D}_0(\theta,\phi)=\left(
	\begin{array}{cccc}
		\frac{1}{81} (3 \cos (2 \theta )-1) & \frac{2}{27} e^{-i \phi } \cos (\theta ) \sin (\theta ) & \frac{2}{27} e^{-i \phi } \cos (\theta ) \sin (\theta ) & \frac{2}{27} e^{-2 i \phi } \sin ^2(\theta ) \\
		\frac{1}{27} e^{i \phi } \sin (2 \theta ) & \frac{1}{81} (1-3 \cos (2 \theta )) & \frac{1}{81} (-3 \cos (2 \theta )-5) & \frac{1}{27} (-2) e^{-i \phi } \cos (\theta ) \sin (\theta ) \\
		\frac{1}{27} e^{i \phi } \sin (2 \theta ) & \frac{1}{81} (-3 \cos (2 \theta )-5) & \frac{1}{81} (1-3 \cos (2 \theta )) & \frac{1}{27} (-2) e^{-i \phi } \cos (\theta ) \sin (\theta ) \\
		\frac{2}{27} e^{2 i \phi } \sin ^2(\theta ) & -\frac{1}{27} e^{i \phi } \sin (2 \theta ) & -\frac{1}{27} e^{i \phi } \sin (2 \theta ) & \frac{1}{81} (3 \cos (2 \theta )-1) \\
	\end{array}
	\right),
\end{equation}
is a traceless matrix.

Finally, the vdW Hamiltonian is thus given by \cref{eq:Hvdw-final} in the main text.

\section{Phonon interactions}
\label{sec:phonon-interactions-app}
In this section, we derive the effective phonon interactions (\cref{eq:phonon-ham-2} in the main text) between two atoms in harmonic traps, separated by a macroscopic distance $r$. 
We assume that the interactions are independent of the internal state and are given by 
\begin{equation}
	\label{eq:Hint-blockade-potential}
	\Hhat_{\text{int}}(r) = \frac{\mathcal{A}}{r^6+R_c^6},
\end{equation} 
where $R_c$ is a blockade radius and $\mathcal{A}$ depends on the vdW interaction strength. 
We further assume that the position of each atom can be decomposed into quantum fluctuations on top of a coherent (classical) part: $\vb{r_i}\rightarrow \vb{r_i} + \vb{\hat{r}_i}$. 
Without loss of generality, we assume that the macroscopic separation $r_0=\abs{\vb{r}_1-\vb{r}_2}$ is along the $y$ direction. In this case, to second order in the small quantum fluctuations we get
\begin{equation}
	\label{eq:Hint-expansion}
	\Hhat_{\text{int}} \approx \Hhat_{x} +\Hhat_y +\Hhat_{z} + \text{Constants},
\end{equation}
where
\begin{equation}
	\label{eq:HxHyHz}
	\begin{split}
		\Hhat_{x} &= - 3\mathcal{A}\frac{(\xhat_1-\xhat_2)^2}{r_0^8\qty[1+(R_c/r_0)^6]^2},\\
		\Hhat_{y} &= -6\mathcal{A}\frac{(\yhat_1- \yhat_2)}{r_0^7\qty[1+(R_c/r_0)^6]^2} 
		+ 3\mathcal{A}\frac{( \yhat_1- \yhat_2)^2\qty[7-5(R_c/r_0)^6]}{r_0^8\qty[1+(R_c/r_0)^6]^3},\\
		\Hhat_{z} &=- 3\mathcal{A}\frac{(\zhat_1-\zhat_2)^2}{r_0^8\qty[1+(R_c/r_0)^6]^2}.
	\end{split}
\end{equation}

The full Hamiltonian of the motional degrees of freedom is 
\begin{equation}
	\label{eq:Hfull}
	\Hhat = \sum_{\alpha=x,y,z}\qty[\sum_{i=1,2}\qty(\frac{\Phat_{i,\alpha}^2}{2M} + \frac{1}{2}M\omega_\alpha^2 \alphahat_i^2) +\Hhat_{\alpha}].
\end{equation}
The full Hamiltonian is therefore a sum of three independent, commuting Hamiltonians for the three directions, which means we can analyze each direction separately.
Note that $\Hhat_x$ and $\Hhat_{y}$ have the same form while $\Hhat_y$ contains a linear term. This linear term, which represents the force between the two atoms, is inherently larger than the quadratic term which gives rise to the phonon-swap terms. 
This fact prevents an efficient cooling of the $y$ direction. In \cref{sec:adiabatic_cooling-app} we show how one can overcome this and nevertheless cool all three directions using an adiabatic protocol.
Here we assume that the confinement along $y$ is sufficiently strong and hence focus on the $x$ and $z$ directions.

The Hamiltonian for the $z$ direction in terms of bosonic creation and annihilation operators $\zhat_1 = \frac{1}{\sqrt{2M\omega_z}}(\ahat_z+\ahat_z^\dagger)$, $\Phat_{1,z}=-i\frac{\sqrt{M\omega_z}}{2}(\ahat_z-\ahat_z^\dagger)$ and $\zhat_2 = \frac{1}{\sqrt{2M\omega_z}}(\dhat_z+\dhat_z^\dagger)$, $\Phat_{2,z}=-i\frac{\sqrt{M\omega_z}}{2}(\dhat_z-\dhat_z^\dagger)$ is given by (the Hamiltonian for $x$ is the same with $z\rightarrow x$)
\begin{dmath}
	\label{eq:phonon-ham-2-SM}
	\Hhat_{\text{ph},z}=\omega_z(\dhat_z^\dagger\dhat_z+\ahat_z^\dagger\ahat_z)-\frac{G_z}{2}\qty[(\dhat_z+\dhat_z^\dagger)^2+(\ahat_z+\ahat_z^\dagger)^2]+G_z(\dhat_z+\dhat_z^\dagger)(\ahat_z+\ahat_z^\dagger),
\end{dmath}
where the phonon-coupling $G_z$ is 
\begin{equation}
	\label{eq:G-def}
	G_z= \frac{3\mathcal{A}}{M\omega_zr_0^{8}}\frac{1}{[1+(R_c/r_0)^6]^2}.
\end{equation}
which is \cref{eq:phonon-ham-2} from the main text, where we dropped the $z$ label. 
Assuming that $\omega_{z}\gg G_z$ and making the rotating wave approximation we have 
\begin{dmath}
	\label{eq:phonon-ham-2-SM-RWA}
	\Hhat_{\text{ph},z}\approx \omega_z(\dhat_z^\dagger\dhat_z+\ahat_z^\dagger\ahat_z)+G_z(\dhat_z\ahat_z^\dagger+\dhat_z^\dagger\ahat_z),
\end{dmath}
or in the rotating frame simply $G_z(\dhat_z\ahat_z^\dagger+\dhat_z^\dagger\ahat_z)$. 

This Hamiltonian effectuates a state-transfer between the two modes $\ahat_z,\dhat_z$, which can be seen from the solution to the Heisenberg equations of motion ($\dot{\dhat}_z(t)=i\comm{G_z(\dhat_z\ahat_z^\dagger+\dhat_z^\dagger\ahat_z)}{\dhat_z},\,\dot{\ahat}_z(t)=i\comm{G_z(\dhat_z\ahat_z^\dagger+\dhat_z^\dagger\ahat_z)}{\ahat_z}$)
\begin{equation}
	\begin{split}
		\ahat_z(t) &= \cos(G_z t)\ahat_z(0)-i\sin(G_zt)\dhat_z(0),\\
		\dhat_z(t) &= -i\sin(G_z t)\ahat_z(0)+\cos(G_zt)\dhat_z(0).
	\end{split}
\end{equation}
After a time of $t_s=\frac{\pi}{2G_z}$, the states of the two modes, and hence the phonon occupations, are swapped. 
If, in addition, we have that $\omega_z=\omega_x$ (and accordingly $G_z=G_x$) then the same swap process would cool both the $x$ and $z$ directions.

Finally, let us comment on the higher-order terms that we neglect in the Taylor expansion.
Each term in the expansion of \cref{eq:Hint-blockade-potential} is smaller than the precedent by the dimensionless factor $\sim \frac{1}{r_0}\sqrt{\frac{1}{M\omega_\alpha}}\, (\alpha=x,y,z)$. For Rubidium atoms separated by $3\,\mum$ in $\omega_z/2\pi = 15\,\khz$ traps (assuming $\omega_z<\omega_{x,y}$) this factor is $\sim 0.03$. Moreover, if we work in the regime where the rotating-wave approximation is valid, i.e $\omega_z \gg G_z$, all the terms that do not conserve the total number of excitations, and in particular all the odd powers in the expansion, can be neglected. Thus, in that regime, the leading order correction to \cref{eq:phonon-ham-2-SM} is smaller by the factor $\sim \qty(\frac{1}{r_0}\sqrt{\frac{1}{M\omega_z}})^2\sim 9\times 10^{-4}$.

\section{Adiabatic phonon-swap}
\label{sec:adiabatic_cooling-app}
In this section, we present an adiabatic protocol for performing the phonon-swap. 
As we have discussed in the previous section, the repulsive force between a pair of atoms prevents the simple phonon-swap from taking place for the trap component parallel to the inter-atomic axis. This manifests itself in the presence of the linear term in the $y$ component of \cref{eq:HxHyHz}. We show here how this can be mitigated by implementing a smooth, slowly varying $\pi/2$ pulse.

This adiabatic protocol can be intuitively understood as follows: we imagine slowly turning on and off the interactions [$\mathcal{A}\rightarrow\mathcal{A}(t)$] such that the atoms adiabatically follow the new equilibrium positions, determined by the total potential, which is the sum of the trap potentials and the interactions. During this time, the phonon-swap can take place, swapping the phonon excitations while the displacements slowly change.

We assume the same setup as in the previous section, where the two atoms, data and auxiliary, are initially (at time $t=0$) separated by some distance $r_0\equiv r(0)=y_2^{eq}(0)-y_1^{eq}(0)$ (determined by the trap separation).  
As we slowly increase $\mathcal{A}(t)$, the equilibrium positions (which at $t=0$ are $y_1^{eq}(0)=0,y_2^{eq}(0)=r_0$) slowly change as well. These equilibrium positions are found by minimizing the full potential at each time $t$, and are given by the solutions to the following equations
\begin{equation}
	\label{eq:y_equilb}
	\begin{split}
		M \omega_y^2 y_1^{eq}(t)-\frac{6 \mathcal{A}(t) [y_1^{eq}(t)-y_2^{eq}(t)]^5}{\left[R_c^6+(y_1^{eq}(t)-y_2^{eq}(t))^6\right]^2}&=0,\\
		\frac{6 \mathcal{A}(t) [y_1^{eq}(t)-y_2^{eq}(t)]^5}{\left[R_c^6+(y_1^{eq}(t)-y_2^{eq}(t))^6\right]^2}+M \omega_y^2 (y_2^{eq}(t)-r_0)&=0.
	\end{split}
\end{equation}
Taylor expanding the potential about those equilibrium positions gives (up to constants)
\begin{dmath}
	\Hhat_{\text{ph},y} = \frac{\Phat_{1,y}^2+\Phat_{2,y}^2}{2M} + \frac{1}{2}M\omega_y^2(\hat{y}_1-y_1^{eq}(t))^2+\frac{1}{2}M\omega_y^2(\hat{y}_2-y_2^{eq}(t))^2- M\omega_yG_y(t)[ \hat{y}_1-y_1^{eq}(t)- (\hat{y}_2-y_2^{eq}(t))]^2,
\end{dmath}
where 
\begin{equation}
	G_y(t) =  -\frac{3\mathcal{A}(t)}{M\omega_y}\frac{7-5(R_b/r(t))^6}{r(t)^8\qty[1+(R_b/r(t))^6]^3},\quad r(t)= y_2^{eq}(t)-y_1^{eq}(t).
\end{equation}
The $x$ and $z$ Hamiltonians are still given by \cref{eq:HxHyHz} and \cref{eq:phonon-ham-2-SM} with the only difference being that $G_z,G_x$ are now time-dependent. In the following we therefore first focus on the $y$ component. 
Note that by expanding about the equilibrium positions, we have implicitly assumed that the process is adiabatic. This assumption can be justified self-consistently, as we show later in this section.

We now transform to the bosonic creation and annihilation operators $\yhat_1=\frac{1}{\sqrt{2M\omega_y}}(\ahat_y+\ahat_y^\dagger)$, $\yhat_2=\frac{1}{\sqrt{2M\omega_y}}(\dhat_y+\dhat_y^\dagger)$, $\Phat_{1,y} = i\sqrt{\frac{M\omega_y}{2}}(\ahat_y^\dagger-\ahat_y)$, $\Phat_{2,y} = i\sqrt{\frac{M\omega_y}{2}}(\dhat_y^\dagger-\dhat_y)$ which gives
\begin{dmath}
	\Hhat_{\text{ph},y} = \omega_y(\ahat_y^\dagger\ahat_y+\dhat_y^\dagger\dhat_y)- \omega_y\alpha_1(t)(\ahat_y+\ahat_y^\dagger)-\omega_y\alpha_2(t)(\dhat_y+\dhat_y^\dagger)-\frac{G_y(t)}{2}[ \ahat_y+\ahat_y^\dagger-2\alpha_1(t)- (\dhat_y+\dhat_y^\dagger-2\alpha_2(t))]^2,
\end{dmath}
where $\alpha_i(t)=\sqrt{\frac{M\omega_y}{2}}y_i^{eq}(t)$. Moving to the adiabatic frame with the displacement operator $\hat{D}(t) = \exp[-\alpha_1(t)(\ahat_y^\dagger-\ahat_y)-\alpha_2(t)(\dhat_y^\dagger-\dhat_y)]$ yields
\begin{dmath}
	\label{eq:H_ph_y_ad}
	\Hhat_{\text{ph,y,ad}} = \hat{D}(t)\Hhat_{\text{ph},y} \hat{D}^\dagger(t) +i\dot{D}(t)D^\dagger(t)=\omega_y(\ahat_y^\dagger\ahat_y+\dhat_y^\dagger\dhat_y) -\frac{G_y(t)}{2}\qty[(\ahat_y^\dagger+\ahat_y)^2+(\dhat_y^\dagger+\dhat_y)^2]+G_y(t)(\ahat_y^\dagger+\ahat_y)(\dhat_y^\dagger+\dhat_y)+i\dot{\alpha}_1(\ahat_y-\ahat_y^\dagger)+i\dot{\alpha}_2(\dhat_y-\dhat_y^\dagger).
\end{dmath}

From \cref{eq:H_ph_y_ad} we can see that the adiabatic Hamiltonian for the $y$ component has a similar structure as the $x$ and $z$ Hamiltonians in the previous section in \cref{eq:phonon-ham-2-SM}, with additional non-adiabatic corrections proportional to $\dot{\alpha}_1$ and $\dot{\alpha}_2$.
The adiabaticity condition is therefore $\omega_y \gg \dot{\alpha}_1,\dot{\alpha}_2$, which together with the condition $\omega_y\gg G_y$ allows us to make the rotating wave approximation, giving
\begin{dmath}
	\label{eq:Hph-y-ad-RWA}
	\Hhat_{\text{ph,y,ad}}\approx \omega_y(\ahat_y^\dagger\ahat_y+\dhat_y^\dagger\dhat_y) +G_y(t)(\ahat_y^\dagger\dhat_y+\ahat_y\dhat_y^\dagger).
\end{dmath}
If $\omega_y \gg \dot{\alpha}_1,\dot{\alpha}_2$, then the atoms follow adiabatically the equilibria of the total potential.
This is in fact the justification for the self-consistent assumption mentioned at the beginning of this section.  

\begin{figure}[tb]
	\centering
	\subfloat[]{{\includegraphics[width=0.46\linewidth]{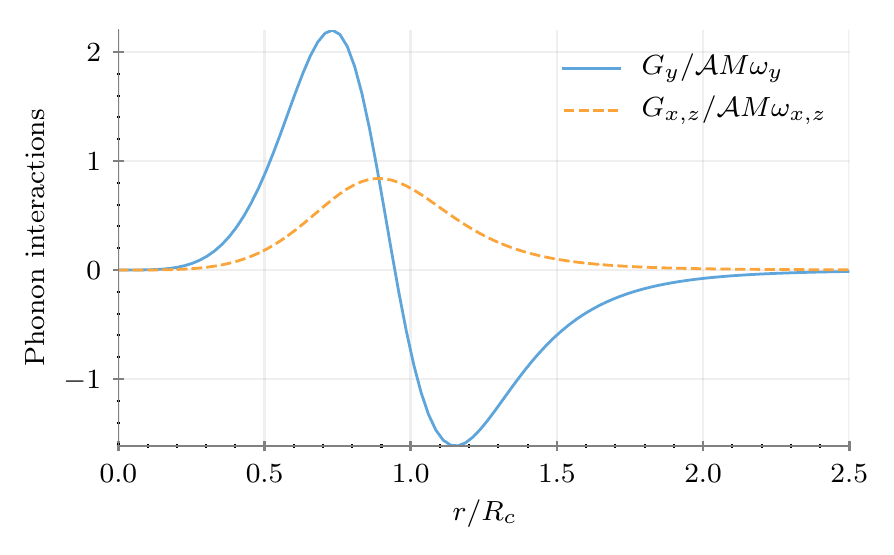} }}%
	\qquad
	\subfloat[]{{\includegraphics[width=0.46\linewidth]{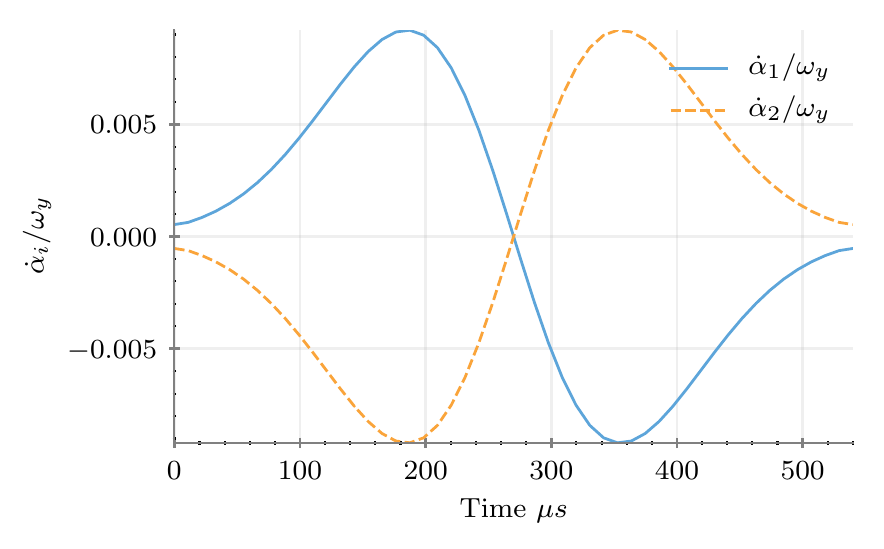} }}%
	\caption{(a) Normalized phonon interactions as a function of distance. (b) Non-adiabatic corrections normalized to the trap frequency $\omega_y/2\pi=50\,\khz$ as a function of time for a Gaussian pulse with $\mathcal{A}_{max}/2\pi =34.4\,\Mhz\,\mum^6,\,\sigma=6.7\,\mics,\,t_0=215.9\,\mics$ for a trap separation of $r_0=1.93\,\mum$. }%
	\label{fig:phonon-int_alphas}%
\end{figure}

\Cref{eq:Hph-y-ad-RWA} effectuates a state-transfer between the two modes, exactly as the time-independent version in the previous section [see \cref{eq:phonon-ham-2-SM-RWA}]. The solution of the Heisenberg equations (in the rotating frame) are in this case (with equivalent expressions for the $x,z$ components)
\begin{equation}
	\begin{split}
		\ahat_y(t) &= \cos(\int G_y(t) dt)\ahat_y(0)-i\sin(\int G_y(t) dt)\dhat_y(0),\\
		\dhat_y(t) &= -i\sin(\int G_y(t) dt)\ahat_y(0)+\cos(\int G_y(t) dt)\dhat_y(0).
	\end{split}
\end{equation}
For a full phonon-swap of the $y$ phonons to take place we require $\int G_y(t) dt=\frac{\pi}{2}$. However, since the phonon interaction strength for the $y$ direction is different than for the $x$ and $z$ directions [i.e., $G_y(t)\neq G_{x,z}(t)$ as one can see in \cref{fig:phonon-int_alphas}(a)] a $\pi/2$ pulse for $y$ is not necessarily a $\pi/2$ pulse for the other two directions. Nevertheless, in typical scenarios the traps are not isotropic, and one can utilize this fact together with the different interaction curves to compensate and optimize a pulse that is as close as possible to $\pi/2$ for all three directions. 

As a simple example, we take a Gaussian pulse for $\mathcal{A}(t)=\mathcal{A}_{max}\qty[\exp(-\frac{(t-t_0)^2}{2\sigma^2})-c]$ where $c$ is a constant chosen such that $\mathcal{A}(0)=0$. Using similar parameters as in the main text (see caption of \cref{fig:phonon-int_alphas}), assuming $\omega_x/2\pi=\omega_z/2\pi=15\,\khz$ and $\omega_y/2\pi=50\,\khz$ we can find pulse parameters that yield $\int G_y(t)dt/(\pi/2) \approx \int G_{x,z}(t)dt/(\pi/2) \approx 1$. 
Furthermore, even if for a given parameter regime it is not possible to perform the 3D phonon-swap with a single pulse, one can always perform the cooling in three steps: one can first apply a $\pi/2$ pulse to swap the phonons in the $x$ and $z$ components. This would only partially swap the $y$ phonons. One then has to cool the auxiliary atoms before applying another $\pi/2$ pulse, this time designed to fully swap the $y$ phonons.

Finally, in \cref{fig:phonon-int_alphas}(b) we show the ratios of $\dot{\alpha}_{1,2}$ to the trap frequency $\omega_y$ for the Gaussian pulse described above. As one can see, the adiabaticity constraint is satisfied well at all times.

\section{Phonon-swap for 1D chain}
\label{sec:1dchain-deriv-app}
Here we generalize the results of the previous sections to the more realistic case of an atomic register consisting of many atoms. We derive \cref{eq:average_n_data} from the main text.

For concreteness we consider the setup in \cref{fig:diagram}(a) of the main text where a 1D chain of $N$ auxiliary atoms (lattice constant $x_0$) is brought to a distance of $y_0$ from an identical 1D chain of $N$ data atoms.
For simplicity we only consider a single trap direction ($z$) and the time-independent swap protocol. The generalizations to two or three directions and the adiabatic swap are straightforward.
The Hamiltonian for the vibrational modes in the $z$ direction is (we drop the $z$ labels from here on)
\begin{dmath}
	\label{eq:H-chain-full}
	\Hhat = \sum_{i=1}^N\omega_z (\ahat_i^\dagger\ahat_i + \dhat_i^\dagger\dhat_i)-\frac{1}{2}\sum_{i=1,j=1,i\neq j}^NG_{ij}\qty[(\zhat_i^a-\zhat_j^a)^2+(\zhat_i^d-\zhat_j^d)^2]-\sum_{i=1, j=1}^NF_{ij}(\zhat_i^a-\zhat_j^d)^2,
\end{dmath}
where by $\ahat_i\,(\dhat_i)$ we denote the phonon-annihilation operator for an auxiliary (data) atom $i$ and by $z_i^a=\frac{1}{\sqrt{2}}(\ahat+\ahat^\dagger) (z_i^d=\frac{1}{\sqrt{2}}(\dhat+\dhat^\dagger))$ the $z$ coordinate of an auxiliary (data) atom (we have absorbed $M\omega_z$ into the definition of $G_{ij}$ and $F_{ij}$).
The $\frac{1}{2}$ is to avoid double-counting and the coefficients are given by
\begin{align}
	\begin{split}
		G_{ij} &=\frac{G}{\eta^{8}\abs{i-j}^{8}},\\
		F_{ij} &=\frac{G}{\qty[\eta^2(i-j)^2+1]^{4}},
	\end{split}
\end{align}
where $G$ is defined in \cref{eq:G-def} with $r=y_0$ and $\eta\equiv\frac{x_0}{y_0}$. To be consistent with the two-atom case, we have defined $G$ with the nearest neighbor separation between data-auxiliary atoms ($y_0$). We have also assumed that pairs of atoms that are farther apart than the nearest-neighbor separation (i.e next nearest-neighbors and so on) experience power-law interactions. In other words, we assumed that the separation between next-nearest neighbors is significantly larger than the blockade radius.
In terms of bosonic operators \cref{eq:H-chain-full} is 
\begin{dmath}
	\Hhat = \sum_{i=1}^N\qty[\tilde{\omega}_{z,i}(\ahat_i^\dagger\ahat_i+\dhat_i^\dagger\dhat_i)-\frac{1}{2}(\tilde{G}_i+\tilde{F}_i)(\ahat_i^2+\dhat_i^2+\text{H.c.})]+\frac{1}{2}\sum_{i\neq j}G_{ij}(\ahat_i\ahat_j+\ahat_i^\dagger\ahat_j+\dhat_i\dhat_j+\dhat_i^\dagger\dhat_j+\text{H.c.})+\sum_{ij}F_{ij}(\ahat_i\dhat_j+\ahat_i\dhat_j^\dagger+\text{H.c.}),
\end{dmath}
where
\begin{align}
	\begin{split}
		\tilde{\omega}_{z,i}&=\omega_{z}+\tilde{G}_i+\tilde{F}_i,\\
		\tilde{G}_{i} &=\sum_{j\neq i}G_{ij},\\
		\tilde{F}_{i} &=\sum_{j}F_{ij}.\\
	\end{split}
\end{align}
We now assume that the system is translationally invariant, i.e, $\tilde{\omega}_{z,i}\approx\tilde{\omega}_{z},\tilde{G}_i\approx\tilde{G},\tilde{F}_i\approx\tilde{F}$ for all $i$. This is a good approximation for the ``bulk" of the atoms, away from the edges, in the limit where $N\rightarrow \infty$, or for a system with periodic boundary conditions. We also assume that $\tilde{\omega}_{z}\gg \tilde{G},\tilde{F}$ which allows us to drop terms that do not conserve the total number of excitations. With these assumptions, the Hamiltonian in the rotating frame is given by
\begin{equation}
	\Hhat=\sum_{i\neq j}G_{ij}(\ahat_i^\dagger\ahat_j+\dhat_i^\dagger\dhat_j)+\sum_{ij}F_{ij}(\ahat_i\dhat_j^\dagger+\ahat_i^\dagger\dhat_j).
\end{equation}

Taking the Fourier transform with
\begin{equation}
	\ahat_n = \frac{1}{\sqrt{N}}\sum_k \ahat_k e^{iknx_0},\qquad \ahat_k =  \frac{1}{\sqrt{N}}\sum_n \ahat_n e^{-iknx_0},
\end{equation}
and using the fact that $G_{ij}$ and $F_{ij}$ are transitionally invariant, i.e depend on $\abs{i-j}$, we get
\begin{equation}
	\label{eq:HkspaceRWA}
	\Hhat=\sum_{k}\qty[G_{k}(\ahat_k^\dagger\ahat_k+\dhat_k^\dagger\dhat_k)+F_{k}(\ahat_k\dhat_k^\dagger+\ahat_k^\dagger\dhat_k)],
\end{equation}
where we used $\sum_ne^{i(k-k')nx_0}=N\delta_{k,k'}$ and the following definitions
\begin{align}
	\begin{split}
		G_{k} &= \sum_{n=-N,n\neq 0}^N G_n e^{-iknx_0}=2\sum_{n=1}^NG_n\cos(knx_0),\\
		F_{k} &= \sum_{n=-N}^N G_n e^{-iknx_0}=F_{n=0}+2\sum_{n=1}^NF_n\cos(knx_0).\\
	\end{split}
\end{align}
\Cref{eq:HkspaceRWA} can be diagonalized with the transformation 
\begin{equation}
	\chat_k = \frac{\ahat_k+\dhat_k}{\sqrt{2}},\qquad \bhat_k = \frac{\ahat_k-\dhat_k}{\sqrt{2}},
\end{equation}
which gives
\begin{equation}
	\Hhat = \sum_{k}\qty[(G_{k}+F_k)\chat_k^\dagger\chat_k+(G_k-F_k)\bhat_k^\dagger\bhat_k].
\end{equation}
Using this, we can now compute the average excitation number in the auxiliary and data atoms, given by
\begin{align}
	\begin{split}
		\bar{n}_a(t) &= \frac{1}{N}\sum_n\expval{\ahat_n^\dagger(t)\ahat_n(t)},\\
		\bar{n}_d(t) &= \frac{1}{N}\sum_n\expval{\dhat_n^\dagger(t)\dhat_n(t)}.\\
	\end{split}
\end{align}
Below, we first compute $\bar{n}_a(t)$:
\begin{dmath}
	\label{eq:nc_long_calc}
	\bar{n}_a(t)= \frac{1}{N}\sum_n\expval{\ahat_n^\dagger(t)\ahat_n(t)} = \frac{1}{4N^2}\sum_k\sum_{nm}e^{ikx_0(n-m)}\expval{4\cos[2](F_kt)\ahat_n^\dagger\ahat_m+4\sin[2](F_kt)\dhat_n^\dagger\dhat_m+2i\sin(2F_kt)\ahat_n^\dagger\dhat_m-2i\sin(2F_kt)\dhat_n^\dagger\ahat_m}.
\end{dmath}
For simplicity, we now assume that the initial state is a product state and also that $\expval{\ahat_i}=\expval{\ahat^2_i}=\expval*{\dhat_i}=\expval*{\dhat^2_i}=0$ for all $i$. This would be the case, for example, if every atom starts at a pure Fock state or a thermal state. With this assumption, only the diagonal terms in \cref{eq:nc_long_calc} contribute, yielding
\begin{equation}
	\bar{n}_a(t) = \frac{1}{N}\qty[\sum_k\cos[2](F_kt)]\bar{n}_a(0)+\frac{1}{N}\qty[\sum_k\sin[2](F_kt)]\bar{n}_d(0).
\end{equation}
Taking the continuum limit $\frac{1}{N}\sum_k\rightarrow \frac{x_0}{2\pi}\int_{-\pi/x_0}^{\pi/x_0}dk$ and changing variables $kx_0\rightarrow k$ gives
\begin{equation}
	\bar{n}_a(t) = \bar{n}_a(0)\int_{-\pi}^{\pi}\frac{dk}{2\pi}\cos[2](F_kt)+\bar{n}_d(0)\int_{-\pi}^{\pi}\frac{dk}{2\pi}\sin[2](F_kt).
\end{equation}

To obtain a closed form expression, we approximate the sum in $F_k$ by the first term $n=1$ which corresponds to only keeping up to next nearest-neighbors interactions between auxiliary and data atoms. This gives rise to 
\begin{align}
	\begin{split}
		\bar{n}_a(t)&=\frac{\bar{n}_a(0)+\bar{n}_d(0)}{2} + \frac{\bar{n}_a(0)-\bar{n}_d(0)}{2}J_0\qty[\frac{4Gt}{(1+\eta^2)^{4}}]\cos(2Gt),\\
		\bar{n}_d(t)&=\frac{\bar{n}_a(0)+\bar{n}_d(0)}{2}  -\frac{\bar{n}_a(0)-\bar{n}_d(0)}{2}J_0\qty[\frac{4Gt}{(1+\eta^2)^{4}}]\cos(2Gt),
	\end{split}
\end{align}
where $J_0(z)$ is a Bessel function of the first kind.

\section{Laser excitation from ground states}
\label{sec:laser-dressing-states-app}

In this section we give an example level structure and laser polarization choice for $^{87}$Rb atoms for some of the schemes we presented in the main text.
One choice for the spin-1/2 states of the data atoms are the following two hyperfine ground states 
\begin{equation}
	\begin{split}
		&\ket{g_-}\equiv|5^2S_{1/2},F=1,m_F=1\rangle,\\
		&\ket{g_+}\equiv|5^2S_{1/2},F=2,m_F=2\rangle.
	\end{split}
\end{equation}
To excite to $S_{1/2}$ states, we need to use an intermediate $P$ state.  Using $\sigma_+,\sigma_-,$ and $\sigma_0$ polarized light, one can for example use the following ladder scheme
\begin{equation}
	\begin{split}
		&|g_-\rangle\xrightarrow{\sigma_0} \ket{5P_{3/2},F=1,m_F=1}\xrightarrow{\sigma_+}\ket{nS_{1/2},m_J={\textstyle+\frac12}},\\
		&|g_+\rangle\xrightarrow{\sigma_0} \ket{5P_{3/2},F=2,m_F=2}\xrightarrow{\sigma_-}\ket{nS_{1/2},m_J={\textstyle-\frac12}}.
	\end{split}
\end{equation}
For the auxiliary atoms, a single state out of the two is sufficient. 
For exciting to $P_{1/2}$ states, one choice is the following
\begin{equation}
	\begin{split}
		&|g_-\rangle\xrightarrow{\sigma_+} |nP_{1/2},m_J={\textstyle+\frac12}\rangle,\\
		&|g_+\rangle\xrightarrow{\sigma_-} |nP_{1/2},m_J={\textstyle-\frac12}\rangle.
	\end{split}
\end{equation}

\end{document}